\documentclass[aps,showpacs,twocolumn,prl]{revtex4}
\usepackage{graphicx}

\begin{document}

\title{A connection between anomalous Poisson-Nernst-Planck models and equivalent circuits with constant--phase elements }
\author{E. K. Lenzi$^1$, J. L. de Paula$^1$, F. R. G. B. Silva$^1$, and L. R. Evangelista$^1$}
\affiliation{$^1$Departamento de F\'{i}sica, Universidade Estadual de Maring\'a,
Avenida Colombo 5790, 87020-900 Maring\'a, Paran\'a, Brazil}

\date{\today}

\begin{abstract}
A connection between the impedance spectroscopy response of anomalous Poisson-Nernst-Planck (PNPA) diffusional models and of equivalent circuits containing constant phase elements (CPE) is established for a typical electrolytic cell. The analysis is carried out in the limit of low frequency in order to highlight the surface effects and to explore how they can be connected to the presence of CPE in the circuit. It is shown that, depending on the choice of the equivalent circuit, the action of these elements can be the same as the one obtained by using integro-differential boundary conditions to describe anomalous diffusive processes in the framework of PNPA models. The predictions are also compared with an experimental data obtained from an electrolytic solution.
\end{abstract}
\maketitle

The AC small--signal immittance (impedance or admittance) spectroscopy (IS) is a powerful method of characterizing many of the electrical properties of materials, and can be used to investigate the dynamics of bound or mobile charge in the bulk or interfacial regions of any kind of liquid or solid material like ionic, semiconducting, mixed electronic-ionic and even insulator (dielectrics) materials~\cite{1}. The models frequently used to analyze the data  are essentially based on continuity equations, for the bulk density of positive and negative ions, satisfying the Poisson's equation requirement for the electric potential across the sample (Poisson--Nernst--Planck or PNP model), or on equivalent circuits with ideal resistors, capacitors, perhaps inductances, and possibly various distributed circuit elements~\cite{1}. A discussion about the various distributed circuit elements that can be incorporated into equivalent circuits was presented in~\cite{2}. However, as pointed out in Ref.~\cite{3}, it is necessary a careful analysis before reaching to general conclusions about the data,  since the incorrect choice of the equivalent circuit can lead to deceptive conclusions about the process that occurs in the cell. Even more powerful and useful general models, such as ordinary  (PNP) or anomalous diffusion (PNPA) ones are not free from ambiguities, as recently discussed by Macdonald~\cite{3b}.

	The  PNPA models aim at incorporating behaviors that may not be well described in terms of usual diffusive PNP models. In electrolytic cells, the anomalous response that generalizes the Warburg model for the electrical impedance was proposed in Ref.~\cite{4}. Bisquert and coworkers~\cite{5,6,7,8,9} have investigated several models with the purpose of determining the electrochemical impedance by using fractional calculus. In this same direction, in Ref.~\cite{10} is evaluated the influence of ions on the IS response of a cell using a complete model in which the fractional drift-diffusion problem is analytically solved satisfying the requirement of the Poisson's equation. In Ref.~\cite{11} is proposed a fractional--type diffusional response for regions of finite length thus leading to an alternative model for the electrical impedance whose form is different from the ones treated in Ref.~\cite{10}.  A comparison between the expressions and responses of alternate anomalous diffusion equations, that were presented in Refs.~\cite{10} and~\cite{11},  was carried out in Ref.~\cite{12}, showing that the anomalous diffusion may play an important role in describing the experimental behavior. These anomalous electrical responses can be found in several systems such as fractal electrodes~\cite{13}, nanostructured iridium oxide~\cite{14}, water~\cite{15}, morphology and ion conductivity of gelatin - LiClO4 films~\cite{16}, and ionic solutions~\cite{1}. Very recently, the model proposed in Ref.~\cite{10} has been extended by incorporating integro-differential terms in the boundary conditions to be satisfied by the solutions of the fundamental equations of the PNP or PNPA models Ref.~\cite{18}.  On the other hand, an important extension used in the framework of equivalent circuits is the CPE, whose presence can be connected to the necessity to describe unusual effects in many solid electrode/electrolyte interfaces. For instance,  Jorcin et al.~\cite{19} have pointed out that the results for a solid electrode/electrolyte interface often reveal a frequency dispersion that cannot be described by simple elements such as resistances, capacitances, inductances or convective diffusion impedance. This behavior can be related to surface disorder and roughness~\cite{20,21} (see, also, Ref.~\cite{25}), electrode porosity~\cite{22}, and to electrode geometry~\cite{23}. To summarize, as stated in Ref.~\cite{24}, a dominant model for describing a capacitance that shows frequency dispersion connected to these situations is just the one considering CPE in equivalent circuits.
	
	Having in mind the importance of these two approaches to analyze the experimental data, the aim of this Letter is to establish, in the low frequency limit, a connection between the predictions of  PNPA models and the ones coming from equivalent circuits with CPE models in the context of the IS response of an electrolytic cell. Actually, it will be shown here that the low frequency behavior obtained by CPE models may be very similar to the ones from PNPA models, if the boundary conditions are suitably represented by integro-differential boundary conditions accounting for unusual diffusive processes.
	
	The discussion starts with a summarized presentation of the fundamental equations of the PNPA model together with the general boundary conditions expressed in terms of an integro-differential equation, along the lines discussed in more details in Refs.~\cite{18,26}. The bulk densities of ions $n_{\alpha}$ ($\alpha =+$ for positive and $\alpha=-$ for negative ones) are governed by the fractional diffusion equation of distributed order:
	
\begin{equation}
\label{old1}
A \frac{\partial}{\partial t} n_{\alpha}(z,t) + B \frac{\partial^{\gamma}}{\partial t^{\gamma}} n_{\alpha}(z,t) = -\frac{\partial}{\partial z} j_{\alpha}(z,t),
\end{equation}                 	
where $\gamma$  is the index of the fractional time derivative defined below,  $A$  is dimensionless and  $B$  has dimensions of $t^{\gamma}$.  Here, $\gamma$  is considered  in the interval $0 < \gamma < 2$  in order to cover sub-diffusive ($\gamma <1$) as well as super-diffusive ($\gamma >1$) situations. The fractional operator used in the model is the Caputo's one,  according to the definition of Podlubny~\cite{33}. The drift-diffusion current density is given by:
 \begin{equation}
\label{old3}
j_{\alpha}(z,t)= -{D}\frac{\partial }{\partial z}n_{\alpha}(z,t)\mp \frac{q {D}}{k_{B}T}n_{\alpha}(z,t)\frac{\partial V(z,t)}{\partial z},
\end{equation}	
where $D$ is the diffusion coefficient for the mobile ions (here assumed as equal for positive and negative ones) of charge $q$, $V(z,t)$  is the effective electric potential across a sample of thickness $d$, with the electrodes placed at the positions $z=\pm d/2$, of a Cartesian reference frame in which $z$  is the axis normal to them, $k_B$  is the Boltzmann constant, and $T$  is the absolute temperature. The potential, present in the drift term of  Eq.~(\ref{old3}), is determined by the Poisson's equation

\begin{equation}
\label{old4}
\frac{\partial^{2}}{\partial z^{2}}V(z,t)=-\frac{q}{\varepsilon}\left[n_{+}(z,t)-n_{-}(z,t)\right],
\end{equation}
in which $\varepsilon$  is the dielectric coefficient of the medium (measured in $\varepsilon_0$ units). The solutions of Eq.~(\ref{old1}) have to satisfy the following boundary condition:

\begin{equation}
\label{old5}
j_{\alpha} \left(\pm \frac{d}{2},t \right) = \pm \int_{-\infty}^t dt'\, \kappa_{\alpha} (t-t')\, \frac{\partial^{\nu}}{\partial t'^{\nu} } n_{\alpha}\left(\pm \frac{d}{2}, t'\right),
\end{equation}
in which the temporal kernel, convoluted with the fractional derivative (of order $\nu$) of the bulk density of charges calculated at the surfaces, may be chosen to describe, as particular cases, many other  physical situations considered elsewhere (see, e.g., Refs.~\cite{18,34,35,36}).  In particular, it may be formally  obtained in the context of  the continuous time random walk \cite{RandomWalk1,reaction1}  if reactive boundary conditions were considered, similarly to what was performed in \cite{reactiveboundary1,reactiveboundary2}. The effective electrical potential coming from Eq.~(\ref{old4}) has to obey the condition $V(\pm d/2, t) = \pm (V_0/2) e^{i \omega t} $   on the electrode surfaces, where $\omega$ is the frequency of the applied potential and $V_0$ its amplitude.

 The set formed of Eqs.~(\ref{old1}) to~(\ref{old5}) represents the mathematical statement of a very general PNPA diffusive model. To obtain analytical solutions for this problem is always a formidable task. However, for the investigation of electrical impedance, one usually assumes that the applied periodic potential has a very small amplitude, which corresponds to the AC small-signal limit. Thus, an exact solution and, consequently, an analytical expression for the electrical impedance (or admittance) can be determined. The details of the calculation can be found elsewhere~\cite{18}. However, it is necessary to underline here that, in this limit, one can assume $n_{\alpha}(z,t) = N + \eta(z) e^{i \omega t} $, with $ N \gg |\eta(z) e^{i \omega t} | $, where $N$   represents the number of  ions per unit volume.  This allows one to assume also that $ V(z,t) = \phi(z) e^{i \omega t}$  to analyze the impedance,  since the stationary state is reached. After performing some calculation, one is able to show that the impedance is~\cite{18}:

\begin{equation}
\label{old6}
{\mathcal{Z}} = \frac{1}{i\omega \varepsilon S \beta^2}\, \frac{M/\lambda^2\beta^2  + d E(i\omega) /(2 D)}{1 + (i \omega)^{\nu-1} {\overline{\kappa}}_{\alpha} (i\omega) \left(1+ i \omega \lambda^2/D \right) M/\lambda^2\beta^2 },
\end{equation}
where $S$  is the electrode area, $M = \tanh(\beta d/2)$,  $\beta^2 = F(i \omega)/D + 1/\lambda^2$, $E(i\omega) = F(i\omega) + i \beta \omega {\overline{\kappa}}_{\alpha} (i\omega) M$, and
${\overline{\kappa}}_{\alpha} (i\omega) = e^{-i \omega t} \int_{-\infty}^t \kappa_{\alpha}(t-t') e^{i \omega t'}\, dt'$.  In Eq.~(\ref{old6}), $\lambda= \sqrt{\varepsilon k_B T/( 2 N q^2 )}$ is the Debye's screening length and
$ F(i\omega) = {\cal{A}}(i\omega\tau)+{\cal{B}}(i\omega\tau)^{\gamma}$.

\begin{figure}
\centering \DeclareGraphicsRule{ps}{eps}{}{*}
\includegraphics*[scale=.25,angle=0]{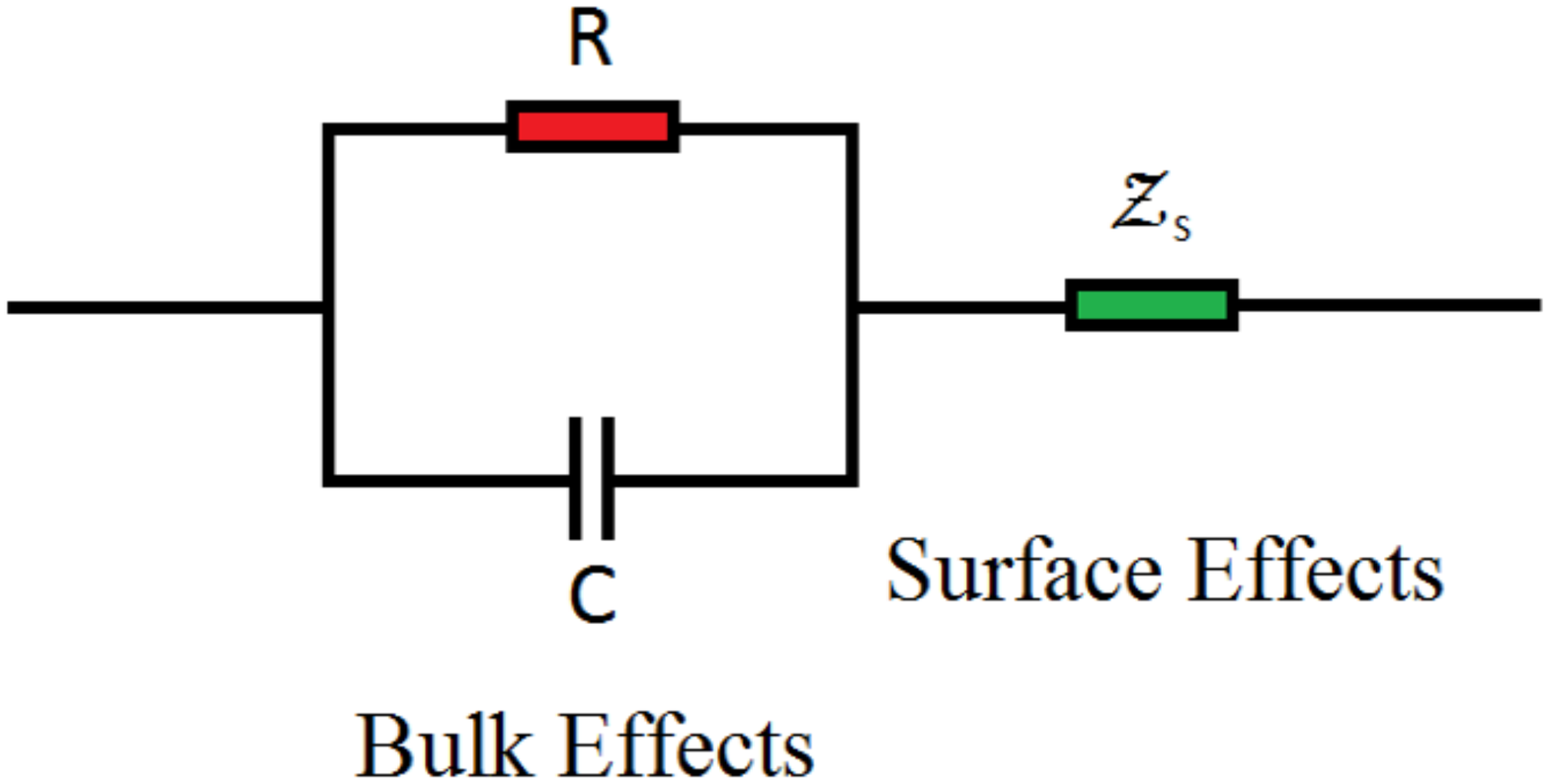}
\caption{Illustration of a circuit in which the first part is a parallel association between a resistive ($R$) and a capacitive ($C$) element. The second part ($\mathcal{Z}_S$) of the circuit is an arbitrary element or association of elements connected to the surface effects.}
\label{FigPNPA1}
\end{figure}

To establish a connection between Eq.~(\ref{old6}) and an equivalent circuit containing a CPE, its low frequency behavior will be determined. In this limit, Eq.~(\ref{old6}) becomes

\begin{equation}
\label{oldnothing}
{\mathcal{Z}}_{\rm PNPA} \approx \frac{2 \lambda^2}{\varepsilon S} \frac{1}{i \omega \left[\lambda + {\overline{\kappa}}_{\alpha} (i\omega)  \right]} + \frac{\lambda^2 d}{\varepsilon S D}
\end{equation}
for $\gamma = \nu =1$. The choice $\gamma =1$ ($B=0$) was made only to simplify the analysis and to connect the bulk effects to a simple association between resistive and capacitive elements, as illustrated in Fig.~\ref{FigPNPA1}. The surface effects are expected to be connected to the second part of the circuit illustrated in Fig.~\ref{FigPNPA1}, i.e., $\mathcal{Z}_S$, which represents an arbitrary element or  an association of elements.  For this reason, an important issue is to know how ${\overline{\kappa}}_{\alpha} (i\omega)$  is connected to $\mathcal{Z}_S$  and, therefore, what is the element that appears when the connection is established.

\begin{figure}
 \centering \DeclareGraphicsRule{ps}{eps}{}{*}
\includegraphics*[scale=.25,angle=0]{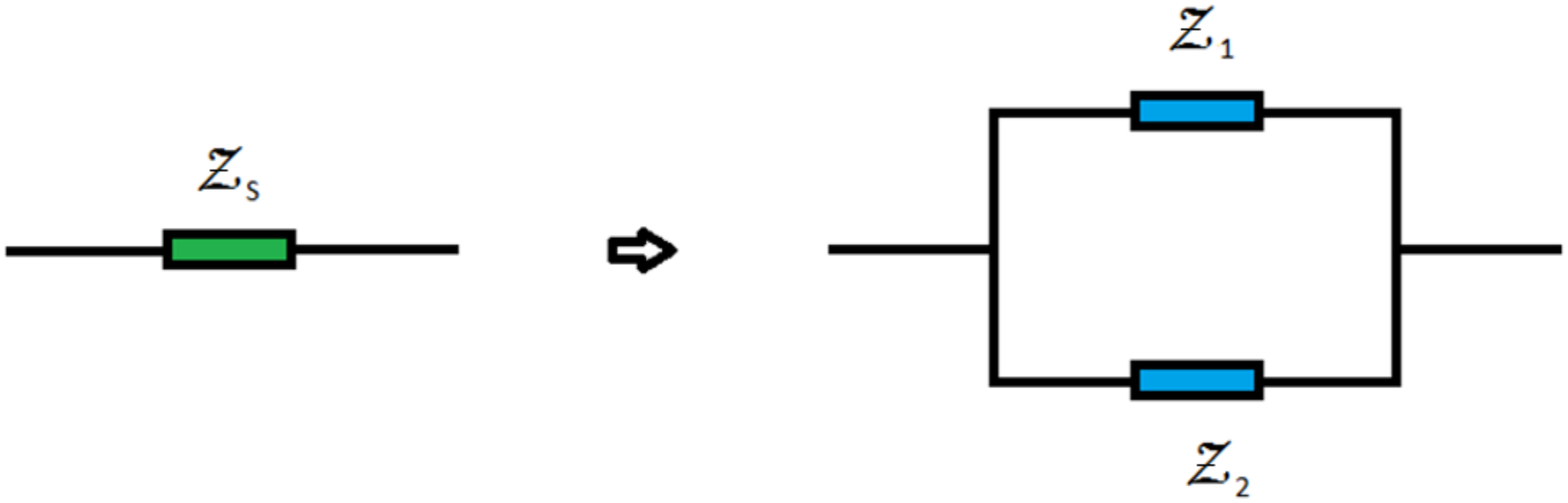}
\caption{Circuit elements forming $\mathcal{Z}_S$ necessary to establish  the connection with the PNPA model in the low frequency limit when ${\overline{\kappa}}_{\alpha} (i\omega)= \kappa \tau/(i \omega \tau)^{\gamma}$.  }
\label{FigPNPA2}
\end{figure}

\begin{figure}
\centering \DeclareGraphicsRule{ps}{eps}{}{*}
\includegraphics*[scale=.25,angle=0]{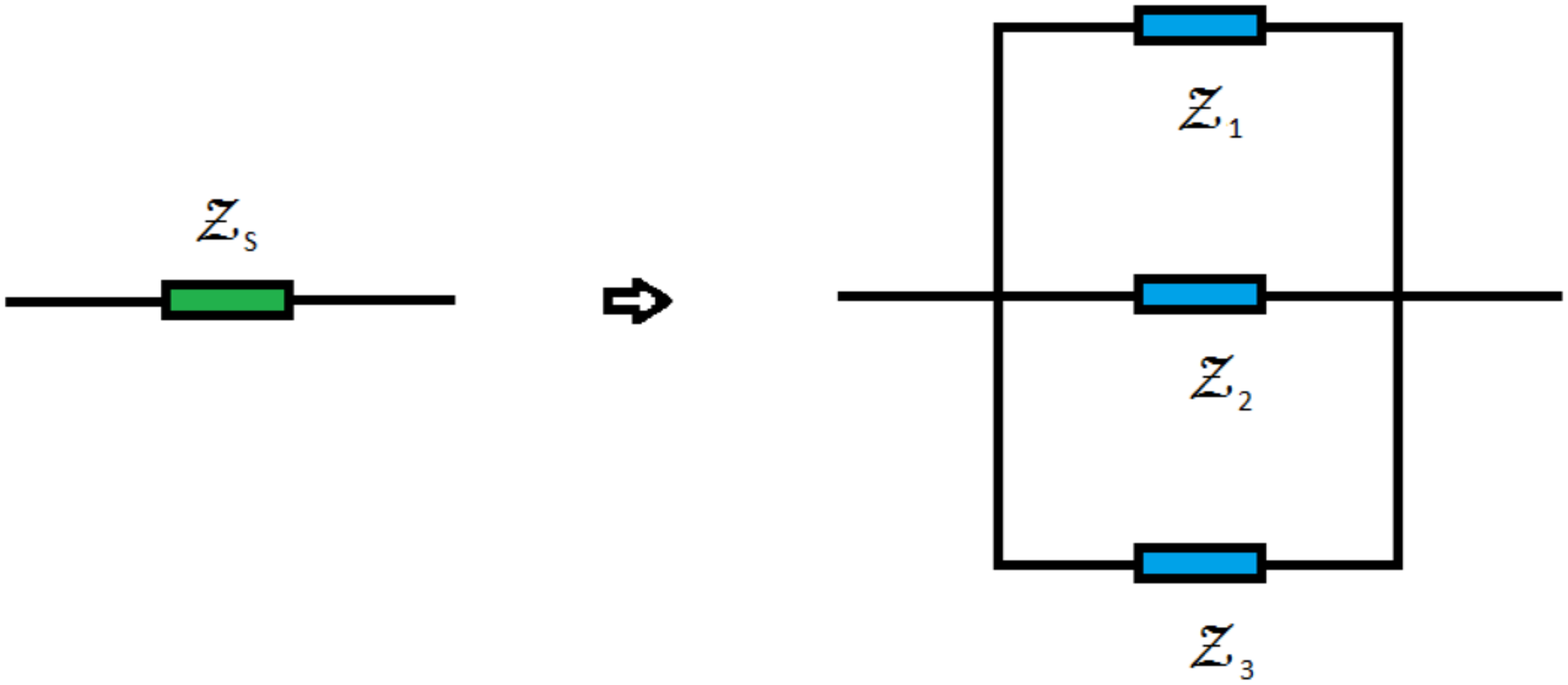}
\caption{Circuit elements forming $\mathcal{Z}_S$ necessary to establish  the connection with the PNPA model in the low frequency limit when ${\overline{\kappa}}_{\alpha} (i\omega)= \kappa_{a,1} \tau_{1}/(i \omega\tau_1)^{\gamma_1} + \kappa_{a,2} \tau_{2}/(i \omega\tau_2)^{\gamma_2}$.  }
\label{FigPNPA3}
\end{figure}

A comparison between ${\mathcal{Z}}_{\rm PNPA}$ and the impedance obtained from the circuit of Fig.~\ref{FigPNPA1},
\begin{equation}
\label{old7}
{\mathcal{Z}_C} = \frac{R}{1+ i \omega R C} + \mathcal{Z}_S,
\end{equation}
in the low frequency limit, yields
\begin{equation}
\label{old8}
\mathcal{Z}_S \approx \frac{2 \lambda^2}{\varepsilon S}\, \frac{1}{i\omega \left[\lambda+{\overline{\kappa}}_{\alpha} (i\omega) \right]}
\end{equation}
where the choice $R = \lambda^2 d/(\varepsilon S D)$  was performed  to relate a bulk effect with the first part of the circuit. Equation~(\ref{old8}) gives a connection between the surface effects represented by ${\overline{\kappa}}_{\alpha} (i\omega)$   and the circuit element or association $\mathcal{Z}_S$. Consequently, for each ${\overline{\kappa}}_{\alpha} (i\omega)$  it is possible to search a simple circuit or an association of circuit elements with the same or equivalent behavior of the impedance,  when the low frequency limit is considered. A typical situation is the one characterized by  perfectly blocking electrodes,  obtained when ${\overline{\kappa}}_{\alpha} (i\omega)=0$, which corresponds to a capacitive element. Other choices for ${\overline{\kappa}}_{\alpha} (i\omega)$  lead to physical processes connected to different surface effects and, therefore, to different elements contributing to $\mathcal{Z}_S$ . Specifically, a relation between the CPE and the boundary conditions used in the PNPA model can be established at this point. To do this, it is useful to rewrite $\mathcal{Z}_S$  as
\begin{equation}
\label{old9}
\frac{1}{\mathcal{Z}_S} \approx \frac{\varepsilon S}{2 \lambda} i \omega + \frac{\varepsilon S}{2 \lambda^{2}} i \omega \overline{\kappa}_{\alpha}(i\omega)
\end{equation}
and to perform  the choice ${\overline{\kappa}}_{\alpha} (i\omega)= \kappa \tau/(i \omega \tau)^{\gamma}$~\cite{18,26}, which, in turn,  implies a parallel association between a capacitor and a CPE. Indeed, $\mathcal{Z}_S$  can be identified with the following association
\begin{equation}
\label{old91}
\frac{1}{\mathcal{Z}_S} \approx \underbrace{\frac{\varepsilon S}{2 \lambda} i \omega}_{1/{\mathcal{Z}_1}} + \underbrace{
\frac{\varepsilon S}{2 \lambda} \frac{\kappa}{\lambda} (i \omega \tau)^{1-\gamma}}_{1/{\mathcal{Z}_2}}
\end{equation}
which represents the association,  exhibited in Fig.~\ref{FigPNPA2},  between a capacitive element, $\mathcal{Z}_1$,  and a CPE,  $\mathcal{Z}_2$, where $\mathcal{Z}_1 = 1/(i \omega C_1)$, with $C_1 = \varepsilon S/(2 \lambda)$,   and $\mathcal{Z}_2 = 1/[(i \omega)^{1-\gamma} C_2]$, with $C_2 = C_1 \kappa \tau^{1-\gamma}/\lambda$.
Other possible choice is to assume ${\overline{\kappa}}_{\alpha} (i\omega)= \kappa_{a,1} \tau_{1}/(i \omega\tau_1)^{\gamma_1} + \kappa_{a,2} \tau_{2}/(i \omega\tau_2)^{\gamma_2}$, which implies

\begin{equation}
\label{old9b}
\frac{1}{\mathcal{Z}_S} \approx \underbrace{\frac{\varepsilon S}{2 \lambda} i \omega}_{1/{\mathcal{Z}_1}} + \underbrace{
\frac{\varepsilon S}{2 \lambda} \frac{\kappa_{a,1}\tau_1}{\lambda} (i \omega \tau_1)^{1-\gamma_1}}_{1/{\mathcal{Z}_2}} + \underbrace{
\frac{\varepsilon S}{2 \lambda} \frac{\kappa_{a,2}\tau_2}{\lambda} (i \omega \tau_2)^{1-\gamma_2}}_{1/{\mathcal{Z}_3}}
\end{equation}
and represents the association illustrated in Fig.~\ref{FigPNPA3}. The elements represented in Fig.~\ref{FigPNPA3}  correspond to a capacitive element ($\mathcal{Z}_1$ ) and two CPE ($\mathcal{Z}_2$ and $\mathcal{Z}_3$), i.e.,  $\mathcal{Z}_1 = 1/(i \omega C_1)$, with $C_1= \varepsilon S/(2 \lambda)$, and $\mathcal{Z}_2 = 1/[(i \omega)^{1-\gamma} C_2]$,
with $C_2= \kappa_{a,1}\tau_1^{1-\gamma_{1}}/\lambda C_{1}$, and $\mathcal{Z}_3 = 1/[(i \omega)^{1-\gamma} C_3]$,  with $C_3= \kappa_{a,2}\tau_2^{1-\gamma_{2}}/\lambda C_{1}$.

\begin{figure}
\centering \DeclareGraphicsRule{ps}{eps}{}{*}
\includegraphics*[scale=.25,angle=0]{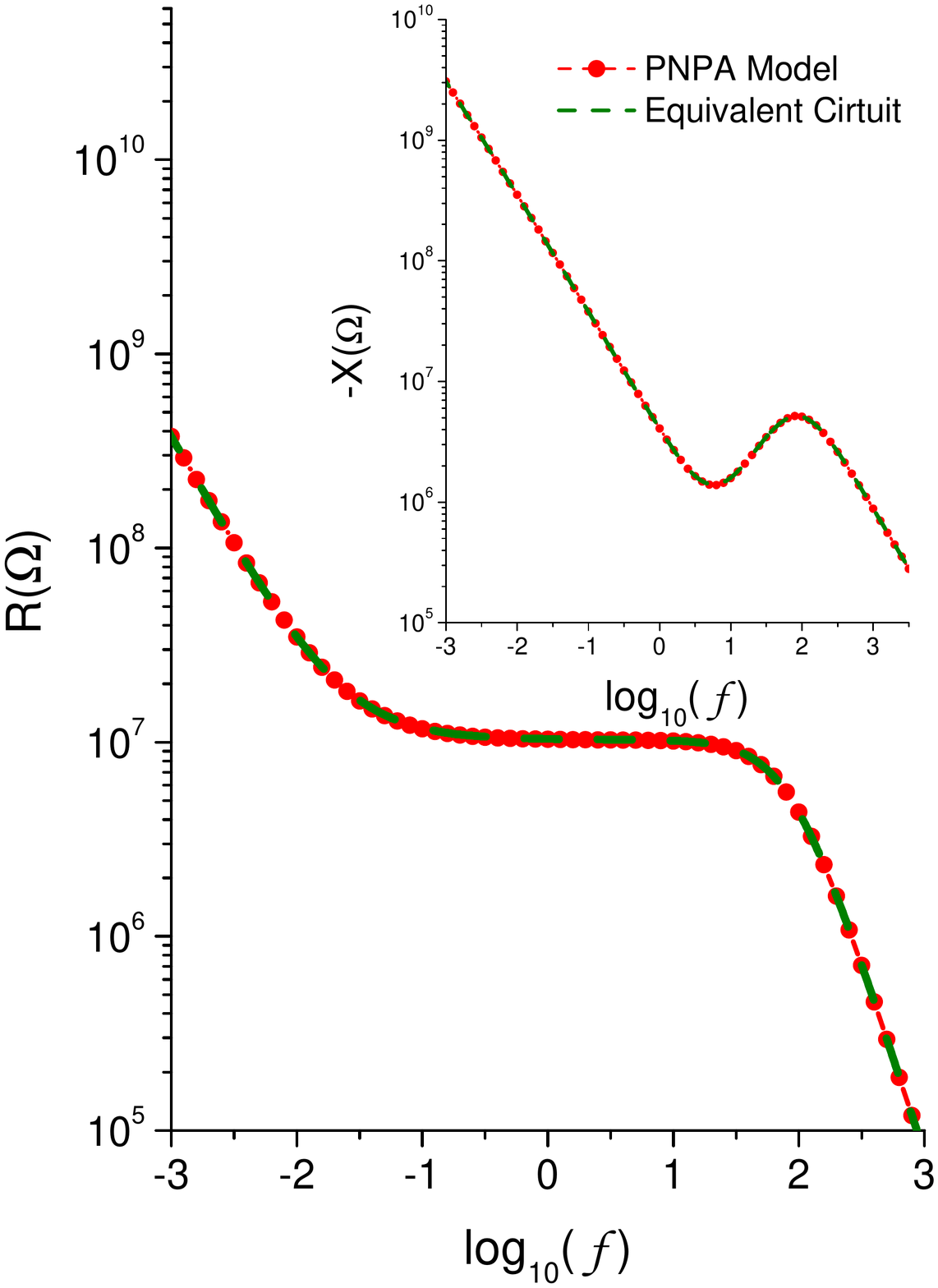}
\caption{Behavior of the real versus imaginary parts of the impedance for the PNPA model and the equivalent circuit arising when Eq.~(\ref{old8}) is used. The red circles represent the PNPA model and the green line is the equivalent circuit obtained from the connection established by Eq.~(\ref{old8}).}
\label{FigPNPA4}
\end{figure}

Figure~\ref{FigPNPA4} illustrates the results for the PNPA model and the equivalent circuit which emerges from the connection established by Equation (\ref{old9}). In this figure,   ${\overline{\kappa}}_{\alpha} (i\omega)= \kappa \tau/(i \omega \tau)^{\gamma}$  and, for simplicity, the parameters values are given in SI units:  $\kappa = 10^{-6}\,$m, $\tau=10^{-3}\,$s, $d= 37\times 10^{-6}\,$m,$\gamma = 0.287$, $\lambda= 8.6 \times 10^{-8}\,$m  $D = 4 \times 10^{-12}\,$m/s, $S = 10^{-4}\,$m$^2$, and $\varepsilon = 7.5 \varepsilon_0$. For the case illustrated here, a good agreement between the PNPA model and the equivalent circuit is obtained when Eq.~(\ref{old8}) is used.

\begin{figure}
\centering \DeclareGraphicsRule{ps}{eps}{}{*}
\includegraphics*[scale=.25,angle=0]{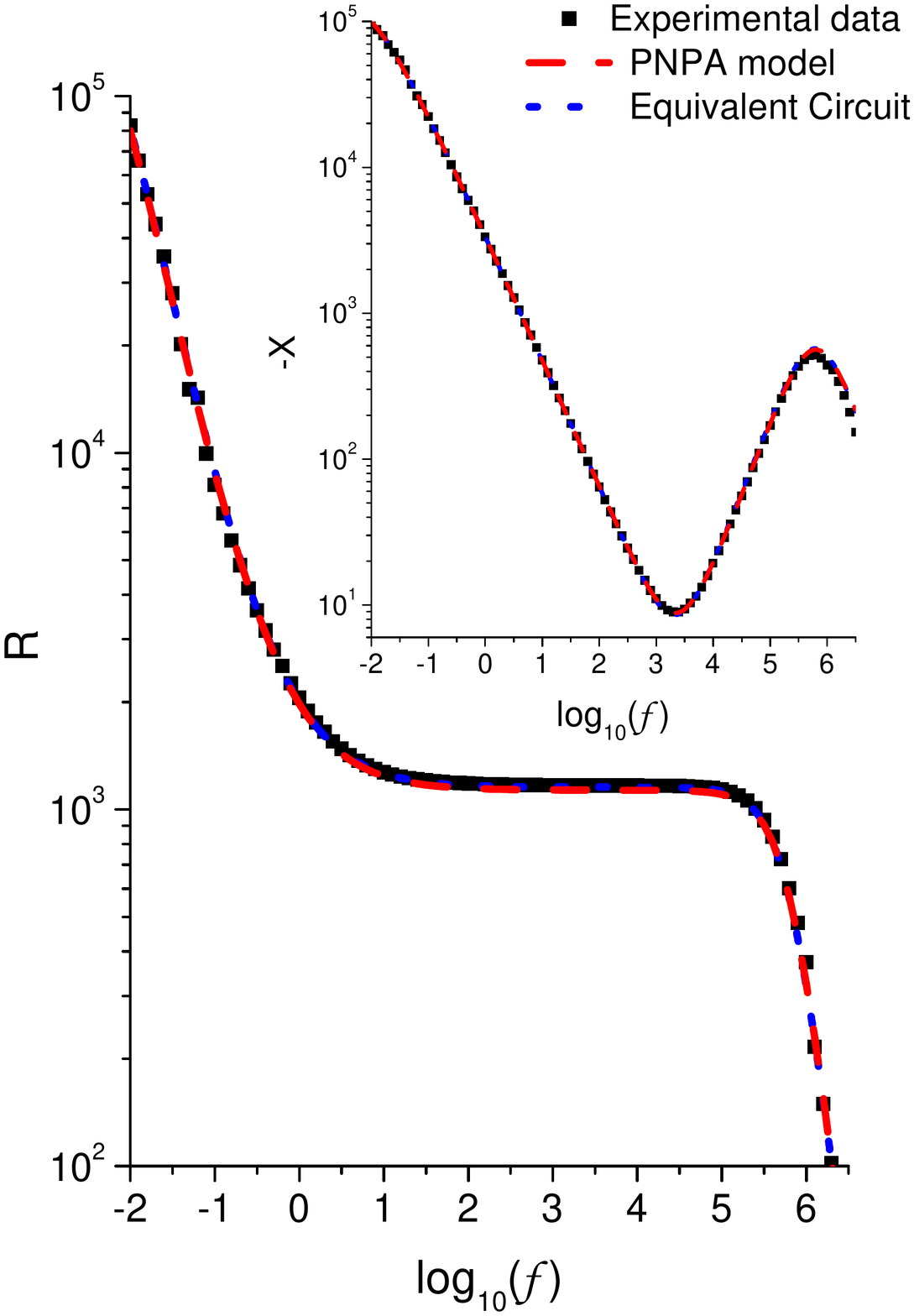}
\caption{ Behavior of the experimental data with the predictions of the model proposed here for the real, $R$, and imaginary, $X$, parts of the impedance. A good agreement between the experimental data and the predictions is obtained  for the parameters: $S=3.14 \times 10^{-4}\;m^2$, $\epsilon=80.03\epsilon_{0}$, $D=3.05 \times 10^{-9}\; m^2/s$, $d=10^{-3}m$, $\kappa_{a,1}=8.67 \times 10^{-5} \;m/s$, $\kappa_{a,2}=6.24\times 10^{-7} m/s$, $\lambda=6.24\times 10^{-8} m$, $\tau=1.64\times 10^{-3}s$, $\gamma_{1}=0.158$, and $\gamma_{2}=0.899$.}
\label{FigPNPA5}
\end{figure}

Let us also consider an experimental scenario to investigate the connection proposed here. The result is illustrated in Fig. \ref{FigPNPA5} which presents the models discussed here and the experimental data of an electrolytic cell of salt (${\mbox{CdCl}}_{2}{\mbox{H}}_2{\mbox{O}}$) dissolved in Milli-Q deionized water (details about the experimental procedure can be found in Ref. \cite{17}). The good agreement, obtained in the context of Fig.~\ref{FigPNPA4}, which compares only the models, is also verified for the frequency range present in Fig.~\ref{FigPNPA5} for the experimental data.
Note that the fit is first obtained between Eq.~(\ref{old6}) and experimental data by using the ``Particle Swarm Optimization'' method~\cite{i53,i54}, where the real and imaginary part of the impedance are simultaneous adjusted with the experimental data. For this case, ${\cal{R}}^2$~\cite{i55,i56} points  out that the model account for about $99.9\%$ of the observed variance in the experimental data. After, we use the parameters, of the model given by Eq.~(\ref{old6}), to obtain the equivalent electric with CPE elements as described above.

In conclusion, a connection between the PNPA models and the whole framework of equivalent circuits with CPE was established on general theoretical grounds. The connection was analytically determined by a careful analysis of the low frequency limit,  where the surface effects play an important role on the electric response of an electrolytic cell. In this limit, we have compared the expressions of the impedance obtained from the PNPA model with the one obtained from an equivalent circuit with an arbitrary component $\mathcal{Z}_S$. This  comparison lead us to a connection between ${\overline{\kappa}}_{\alpha} (i\omega)$  and $\mathcal{Z}_S$ i.e, to the proposition of an equivalence between these two approaches which is very clear in the limit of low frequency but may be also valid in a broader frequency range, as illustrated in Fig.~\ref{FigPNPA4}. However,  in the high frequency limit these approaches may lead to different results depending on the choice of  $\mathcal{Z}_S$. The results presented here permits one to conclude, on analytical grounds, that the effect of a CPE in an equivalent circuit may be represented by an appropriated term in the boundary condition of a PNP or PNPA model. In this regard, the analysis may be helpful to shed some light on the possible meaning of a frequency-domain CPE in terms of a condition formulated in the time-domain at the electrodes limiting the system and offers two conceptual routes to face the complex richness of the impedance spectroscopy data.

\acknowledgments

This work was partially supported by the National Institutes of Science and Technology of Complex Fluids -- INCT-FCx (L. R. E.) and Complex Systems -- INCT-SC (E. K. L.) and Brazilian Agencies Capes (F. R. G. B. Silva) and CNPq (J. L. de Paula).

\end{document}